\documentclass{emulateapj}

\slugcomment{}
\shortauthors{Kitayama et al.}
\shorttitle{Constraints on the Intracluster Dust Emission in the Coma
Cluster}
\begin{document}
%
\title{Constraints on the Intracluster Dust Emission in the Coma
Cluster of Galaxies}

\author{
Tetsu Kitayama\altaffilmark{1}, 
Yuichi Ito\altaffilmark{1}, 
Yoko Okada\altaffilmark{2,8}, 
Hidehiro Kaneda\altaffilmark{2,3}, 
Hidenori Takahashi\altaffilmark{4}, 
Naomi Ota\altaffilmark{2,5}, 
Takashi Onaka\altaffilmark{6},
Yuka Y. Tajiri\altaffilmark{6,7}, 
Hirohisa Nagata\altaffilmark{2}and
Kenkichi Yamada\altaffilmark{1}}
\altaffiltext{1}{Department of Physics, Toho University,  Funabashi,
  Chiba 274-8510, Japan}
\altaffiltext{2}{Institute of Space and Astronautical Science, Japan Aerospace Exploration Agency, 3-1-1 Yoshinodai, Sagamihara, Kanagawa 229-8510, Japan}
\altaffiltext{3}{Division of Particle and Astrophysical Sciences, 
Nagoya University, Chikusa-ku, Nagoya 464-8602, Japan}
\altaffiltext{4}{Gunma Astronomical Observatory, 6860-86 Nakayama,
  Takayama-mura, Agatsuma-gun, Gunma 377-0702, Japan}
\altaffiltext{5}{Max-Planck-Institut f\"ur extraterrestrische Physik, 
85748 Garching, Germany}
\altaffiltext{6}{Department of Astronomy, The University of Tokyo, 
Tokyo 113-0033, Japan}
\altaffiltext{7}{Nippon System Development Co. Ltd., Tokyo, Japan}
\altaffiltext{8}{Present address: I. Physikalisches Institut, 
 Universit\"{a}t zu
      K\"{o}ln, Z\"{u}lpicher Str. 77, 50937 K\"{o}ln, Germany}

\begin{abstract}
We have undertaken a search for the infrared emission from the
intracluster dust in the Coma cluster of galaxies by the Multiband
Imaging Photometer for {\it Spitzer}. Our observations yield the deepest
mid and far-infrared images of a galaxy cluster ever achieved. In each
of the three bands, we have not detected a signature of the central
excess component in contrast to the previous report on the detection by
{\it Infrared Space Observatory (ISO)}. We still find that the
brightness ratio between 70$\mu$m and 160$\mu$m shows a marginal sign of
the central excess, in qualitative agreement with the {\it ISO} result.
Our analysis suggests that the excess ratio is more likely due to faint
infrared sources lying on fluctuating cirrus foreground.  Our
observations yield the 2$\sigma$ upper limits on the excess emission
within 100 kpc of the cluster center as $5 \times10^{-3}$ MJy/sr, $6
\times10^{-2}$ MJy/sr, and $7\times 10^{-2}$ MJy/sr, at 24, 70, and 160
$\mu$m, respectively. These values are in agreement with those found in
other galaxy clusters and suggest that dust is deficient near the
cluster center by more than 3 orders of magnitude compared to the
interstellar medium.
\end{abstract}

\keywords{galaxies: clusters: general -- galaxies: clusters: individual
(Abell 1656) -- infrared: general -- intergalactic medium}

\section{Introduction \label{sec:intro}}

The presence of dust grains in the diffuse intergalactic medium is still
an open question. If they exist, they should have great impacts not only
on our understanding of galaxy evolution but also on interpretation of
high-redshift observational data.  From X-ray measurements of heavy
metal lines, it is evident that clusters and groups of galaxies contain
a significant amount of elements processed in galaxies \citep[e.g.,][and
references therein]{Tamura04}. Such elements are likely to have been
expelled out of galaxies through ram pressure stripping, galaxy mergers
and interactions, blowout by supernovae, etc. Dust may also be expelled
by these processes or by some independent mechanism such as radiation
pressure \citep[e.g.,][]{Chiao72, Ferrara91}.  The X-ray emitting hot
gas can collisionally heat the dust \citep{Dwek86}, while it can also
destroy small dust grains via sputtering \citep{Draine79, Tsai95}. Dust
grains coexistent with the X-ray emitting gas have indeed been detected
in a number of elliptical galaxies
\citep[e.g.,][]{Knapp89,Kaneda07,Temi07}.  The temperature, the amount,
and the size distribution of the dust in hot plasma can be quite
different from those in the normal interstellar medium (see Dwek \&
Arendt 1992 for review on dust--gas interactions).  Once detected firmly,
the intracluster or intragroup dust will definitely provide a powerful
probe of galaxy evolution and the dust--gas interaction.

There have been several suggestions and debates regarding the presence
of dust in clusters and groups of galaxies. Over five decades ago,
\citet{Zwicky57} first suggested an enhanced extinction toward the Coma
cluster of galaxies. More recently, \citet{Girardi92} pointed out that
the redshift asymmetries of member galaxies in nearby groups are
consistent with the presence of dust in the intragroup medium.  The
observed oxygen K edge in an X-ray spectrum of the Perseus cluster can
also be attributed to intracluster dust grains \citep{Arnaud98}.  The
extended submillimeter emission detected in a rich galaxy cluster may
partly be due to dust \citep{Komatsu99, Kitayama04}.  There have
been extensive searches for the enhanced visual extinction toward high
redshift objects behind clusters \citep{Maoz95, Nollenberg03, Muller08} and
\citet{Chelouche07} reported the detection of reddening toward $z\sim 0.2$
clusters.  On the other hand, the intracluster dust is likely to
enhance the efficiency of gas cooling \citep{Dwek87} that is in apparent
contradiction with recent X-ray observations \citep{Peterson03}.

More direct evidence has been searched for in the far-infrared
bands. \citet{Hickson89} reported that the far-infrared emission
observed with {\it IRAS} is enhanced by about a factor of 2 in compact
groups of galaxies compared with a sample of isolated galaxies. The
extended excess emission is also inferred toward Abell 262 and Abell
2670 \citep{Wise93}.  \citet{Sulentic93}, however, pointed out that the
results of \citet{Hickson89} are likely to have been overestimated
mainly due to the limited spatial resolution of {\it IRAS}.
\citet{Stickel98} and \citet{Stickel02} reported the detection of
extended excess far-infrared emission from the Coma cluster, which can
be attributed to thermal emission from the intracluster dust. They find
such emission only in one out of six clusters they observed with {\it
Infrared Space Observatory (ISO)}. \citet{Montier05} combined the {\it
IRAS} maps toward 11,507 galaxy clusters and statistically detected the
excess emission, although the contribution of member galaxies is yet to
be fully removed.

Major difficulties in the previous infrared searches are the limited
sensitivity and spatial resolution of the detectors. These limitations
may have led to significant contamination of individual galaxies as well
as Galactic cirrus confusion. It will therefore be meaningful to perform
infrared observations using current facilities with much improved
sensitivity and spatial resolution. \citet{Bai07} searched for the
intracluster dust emission from Abell 2029 with {\it Spitzer} and
reported that the cirrus noise limits the detection of the intracluster
dust component.

In this paper, we present the results of our deep observations toward
the central region of the Coma cluster (Abell 1656) with the Multiband
Imaging Photometer for {\it Spitzer} (MIPS) on board the {\it Spitzer}
Space Telescope. Coma lies at high galactic latitude ($b=88^\circ$)
where the cirrus level is expected to be among the lowest.  In fact,
\citet{Stickel98} and \citet{Stickel02} found with {\it ISO} the excess
far-infrared emission toward Coma based on the brightness ratio
$I_{120\mu\rm{m}}/I_{180\mu\rm{m}}$, while the S/N at each wavelength is
still insufficient to detect the excess directly and the contamination
of galaxies is not removed. We attempt to probe the excess emission more
directly by approaching the confusion limits of extragalactic sources
and Galactic cirrus in each band.

Throughout the paper, we assume a standard set of cosmological
parameters, $\Omega_m=0.27$, $\Omega_\Lambda=0.73$, and $h=0.71$
consistent with the {\it WMAP} five year data \citep{Komatsu08}.  An
angular size of 1$''$ corresponds to a physical size of 0.46 kpc at the
redshift $z=0.0231$ of the Coma cluster.

\section{Observations and Data Reduction \label{sec:obs}}

\begin{figure}[b]
\epsscale{.95} \plotone{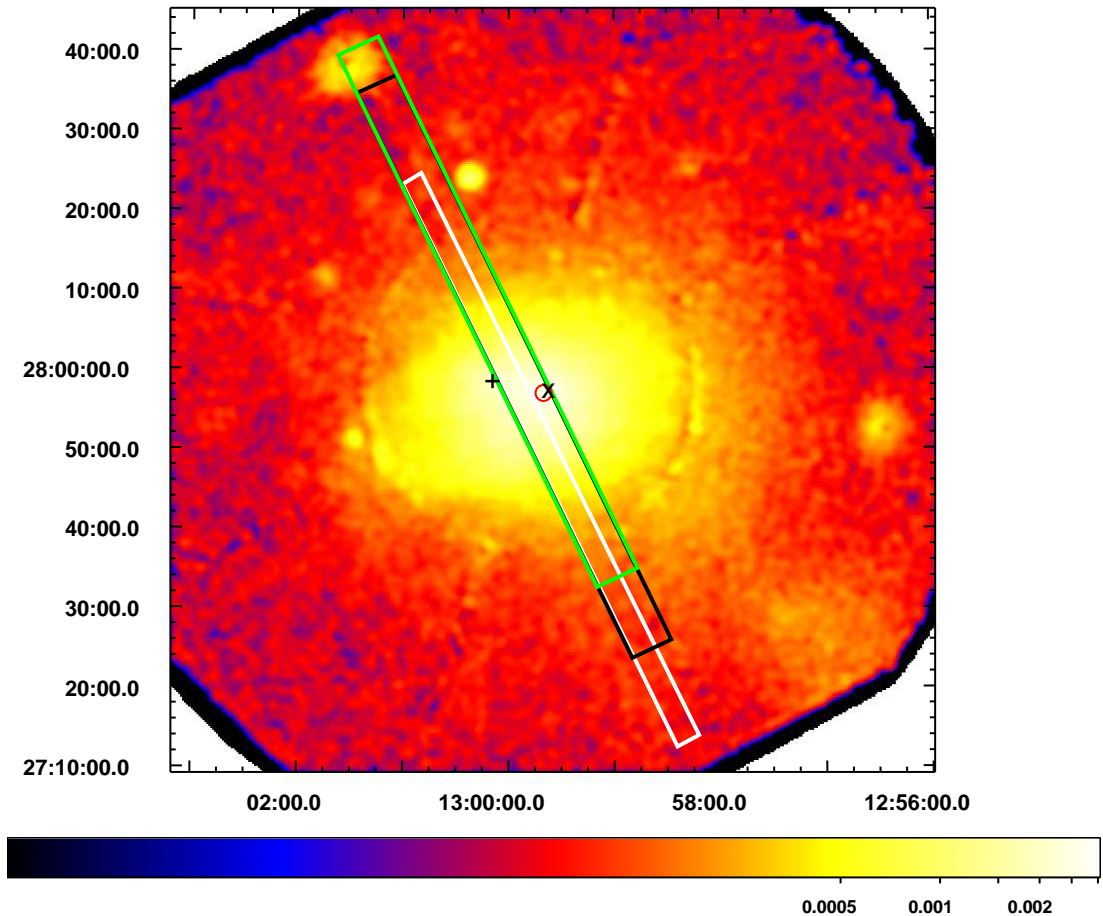} \caption{Fields-of-view of our
     {\it Spitzer}/MIPS observations at 24 $\mu$m (black box), 70 $\mu$m
     (white box) and 160 $\mu$m (green box) overlaid on the {\it ROSAT}
     PSPC X-ray image of the Coma cluster. The X-ray image is scaled
     logarithmically in units of photons/sec/pixel in the 0.2--2 keV
     band, after smoothing by a Gaussian filter with $\sigma=4$ pixel,
     where the pixel size is 14.9 arcsec.  The center of the X-ray
     surface brightness is indicated by a circle, and the positions of
     two massive central galaxies, NGC4874 and NGC4889, are marked by a
     cross and a plus, respectively.}\label{fig:pspc}
\end{figure}

The central region of the Coma cluster was observed at 24, 70 and 160
$\mu$m with the MIPS in our GO3 program on 2007 January 19. Total
duration of the observation was 41 ksec. Linear scans were performed in
a slow scan mode covering the $5'\times 50'$ region across the cluster
as shown in Figure \ref{fig:pspc} (the scan width was reduced to $\sim
2.5'$ at 70 $\mu$m due to dead pixels).  The scan paths were carefully
chosen so as to avoid bright galaxies and to lie close to the cluster
center; the scan center was fixed at the mid-point ($12^{\rm h}59^{\rm
m}51.9^{\rm s}, +27^\circ58'05''$) between two massive central galaxies,
NGC4874 and NGC4889 with the position angles of $-25.4^\circ$,
$-28.9^\circ$ and $-25.2^\circ$ at 24, 70 and 160 $\mu$m, respectively.
The above scan center is $3.2'$ northeast of the X-ray center of the
cluster in the {\it ROSAT} PSPC image ($12^{\rm h}59^{\rm m}40.1^{\rm
s}, +27^\circ56'56''$). This separation is small compared to the extent
of the X-ray emission, which is characterized by the core radius of
$10'$.  In total, 32 scans were performed in each band, 16 of which
repeatedly covered a $5'\times 30'$ region on either side of the scan
center.  The integration time per pixel after 16 scans was 1600, 1600
and 160 sec at 24, 70 and 160 $\mu$m, respectively.  Two sets of the 16
scan legs overlapped and the integration time was doubled in the central
$5'\times 10'$ region, where the X-ray emission is the most prominent.

The data were first processed with the MIPS pipeline, version S17
\citep{Gordon05} to yield the Basic Calibrated Data (BCD).  To further
improve the images, the pixel-dependent and scan-dependent backgrounds
were corrected as follows. In doing so, special care was taken to
maintain the extended nonlinear component of the emission. All the
results regarding the extended emission are based on unfiltered images.

First, we create a median flat field of detector pixels by stacking the
normalized BCD frames, excluding the regions within $10'$ from the
cluster center. The normalization is done by dividing each BCD by its
median. This is intended to separate any large-scale variations, e.g.,
due to the zodiacal light, from the pixel-dependent backgrounds and
applied solely to create the flat field. We have checked that the
flat field so obtained is properly normalized; its mean is equal to
unity within 0.2\%.  We then divide the original BCD (not normalized) by
the normalized median flat field. At 24 $\mu$m, we further remove
scattered light that produces a background modulation as a function of
the scan mirror position; the average offset at each scan mirror
position is computed excluding the data within $10'$ from the cluster
center.  As mentioned below, these processes will be re-run after the
source identification.

Second, we create for each band a mosaic image using the MOPEX software
package, version 18.2 \citep{Makovoz05} developed at the Spitzer Science
Center. The pixel size is taken to be a nominal value of $2.45''$,
$4.0''$ and $8.0''$ at 24, 70 and 160 $\mu$m, respectively. We exclude
the first five BCD frames in each scan leg to omit boost frame
transients. We also exclude the frames affected by stimulator flashes at
70 and 160 $\mu$m.  The background matching is applied {\it after}
combining the frames in each scan leg; we only correct the constant
offsets among 32 scan legs to assure sufficiently large overlapping area
and to preserve any extended feature of the emission.

Third, we identify point sources in the mosaic images down to the
1$\sigma$ detection limit ($\sim 0.02$, $\sim 2$ and $\sim 20$ mJy at
24, 70 and 160 $\mu$m, respectively) 
using APEX \citep{Makovoz05b}. This limit corresponds to the $2 \sim 3
\sigma$ confusion level in each band \citep{Dole04b}. The regions around
the sources are masked out and unused in our analysis. More
specifically, the source identification and the masking procedure are
done on the basis of median-subtracted images; a median-filtered image
is created and subtracted from the unfiltered image solely for this
purpose. To improve the efficiency of source detection against unknown
background fluctuations, we vary the size of a median filter box from $5
\times 5$ to $25 \times 25$ pixels. For every source identified in each
median-subtracted image, we estimate the radius $r_{\rm mask}$ at which
the surface brightness, including that of the side-lobes, drops below a
certain threshold noise level using the Point Response Function. We
adopt 0.5$\sigma$ as the threshold noise level at 24 and 70$\mu$m, and
1$\sigma$ at 160 $\mu$m, considering the lower uniformity of the data in
the latter band. We then return to the {\it unfiltered} image and mask
all the pixels within $r_{\rm mask}$ from the sources identified in any
of the median-subtracted images.  We have checked that the correlation
coefficient between the mean brightness of masked pixels ($r <r_{\rm
mask}$) and that of unmasked pixels at $r_{\rm mask}< r < 2 r_{\rm
mask}$ is less than 0.2 in any of the median-subtracted images. We thus
consider that our masking procedure successfully removes the
contribution of resolved sources.

Strictly speaking, the source identification and the background
correction are not independent of each other. We therefore repeat the
above procedure from the flat-field correction excluding the pixels
around the point sources in the BCD.

After applying the above corrections, we subtract from each unfiltered
image the linear gradient over the entire map, so as to remove the
variation on scales larger than the galaxy cluster. Such a gradient is
greatest at 24 $\mu$m due to the zodiacal light; it causes the
difference of $\Delta I_\nu > 1$ MJy/sr between the map edges which is
more than 2 orders of magnitude larger than the average noise level of
the image (Table \ref{tab:images}).  By simply subtracting the linear
component, we avoid relying on specific models of the zodiacal light
emission that are still uncertain at scales probed here.  To preserve
the average brightness of the entire map, the brightness at the scan
center is kept unchanged by the subtraction.

\section{Results \label{sec:results}}

\begin{figure}
\epsscale{1.1} \plotone{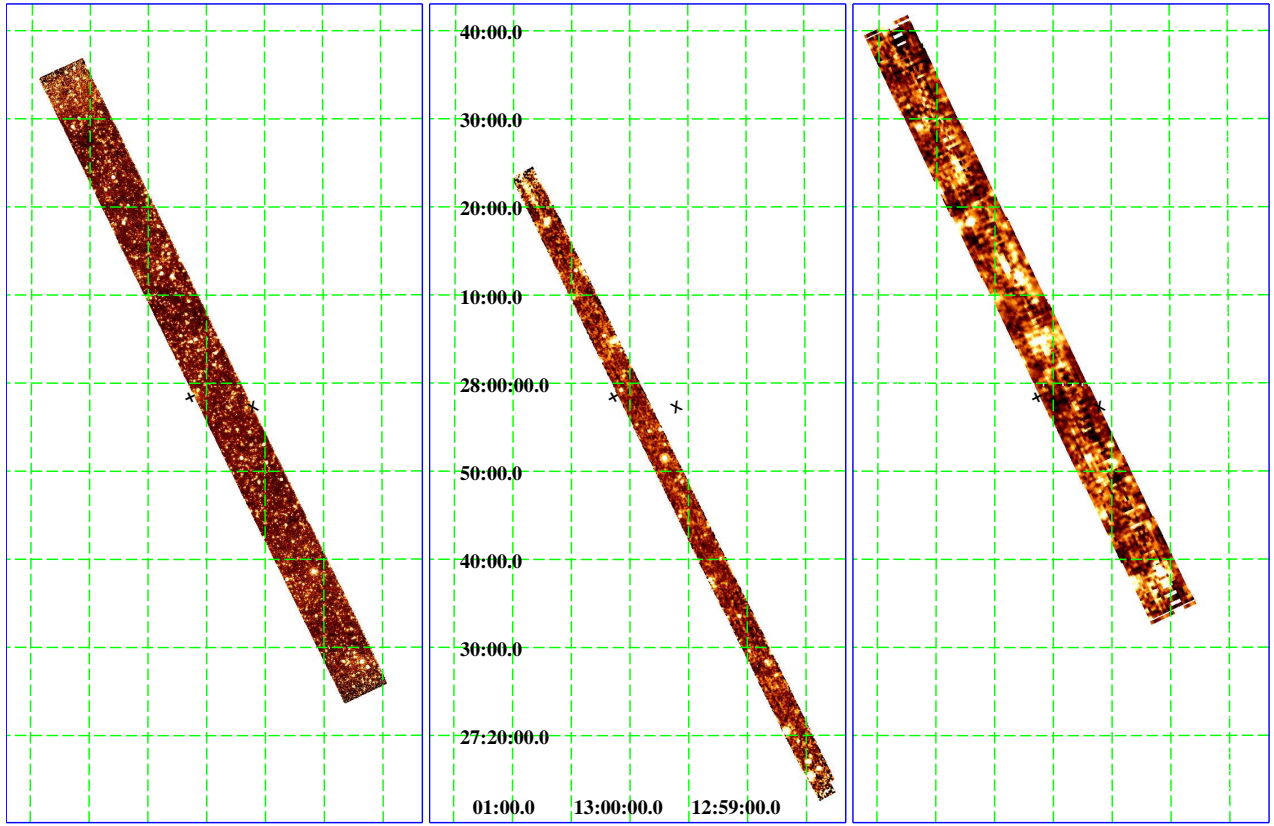} \caption{Mosaic images at 24
$\mu$m ({\it left panel}), 70 $\mu$m ({\it middle panel}), and 160 $\mu$m ({\it
right panel}), respectively. The positions of two massive central galaxies,
NGC4874 and NGC4889, are marked by a cross and a plus, respectively. }
\label{fig:images}
\end{figure}

Figure \ref{fig:images} shows the mosaic images after the subtraction of
the linear gradient. The point sources are explicitly shown for display
purposes. The pixel sizes of the images, the FWHMs of the Point Response
Function, and the average noise levels are listed in Table
\ref{tab:images}. We have assigned a noise to an individual pixel as the
larger of the uncertainty estimated in the MIPS pipeline and the
standard deviation of actual measurements at that pixel. The latter is
on average twice as large as the former at 70 and 160 $\mu$m, and both
are comparable at 24 $\mu$m.  This choice should enable us to take
account of any residual systematics in each pixel more properly. Note
that the sensitivity at 160 $\mu$m is likely limited by Galactic cirrus;
the $1\sigma$ fluctuation level expected at the position of Coma is
$5.4\times 10^{-3}$ MJy/sr at 70 $\mu$m and 0.20 MJy/sr at 160 $\mu$m
\citep{Jeong05}.

\begin{figure}
\epsscale{0.37} \plotone{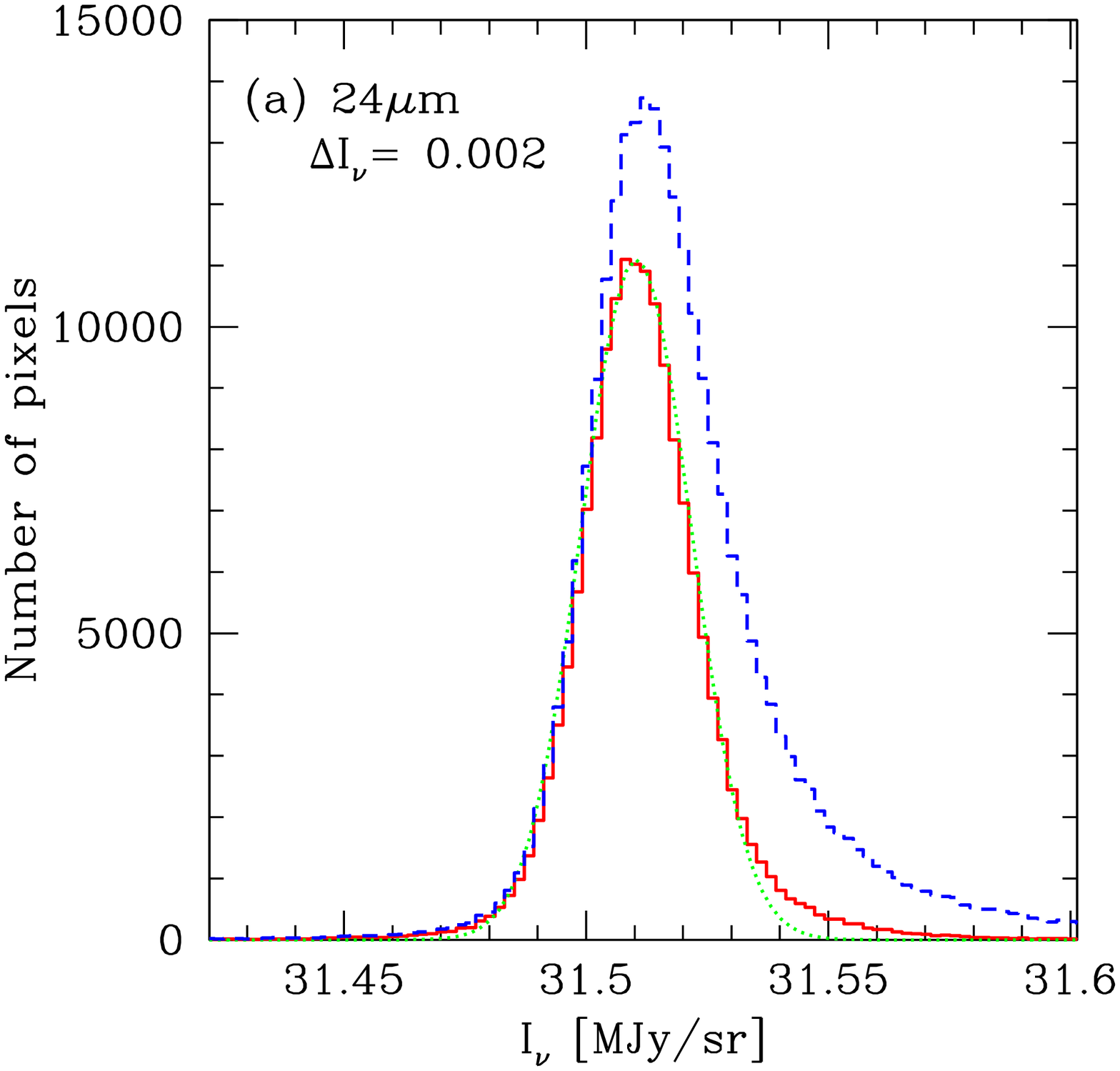} \plotone{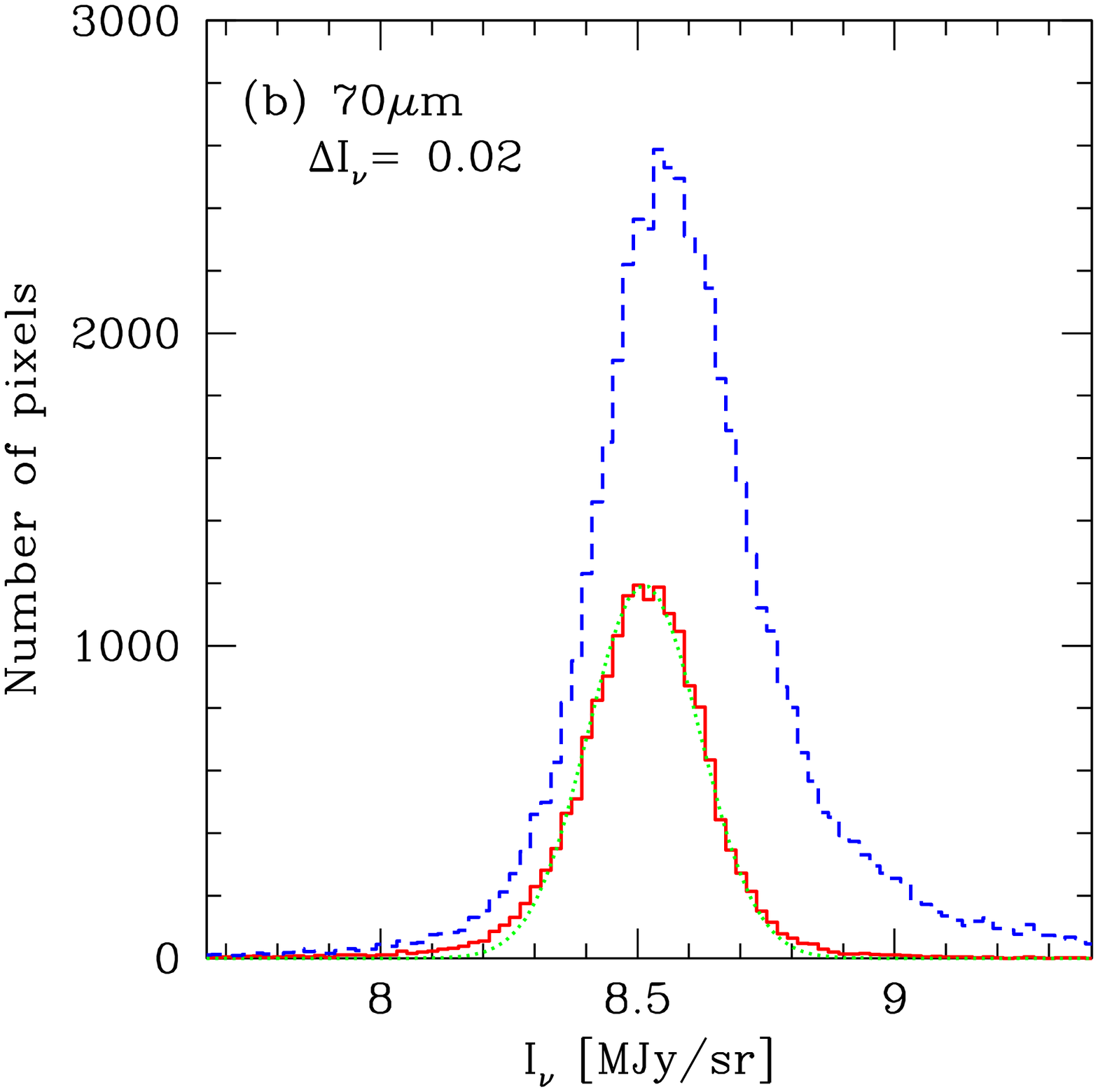}\plotone{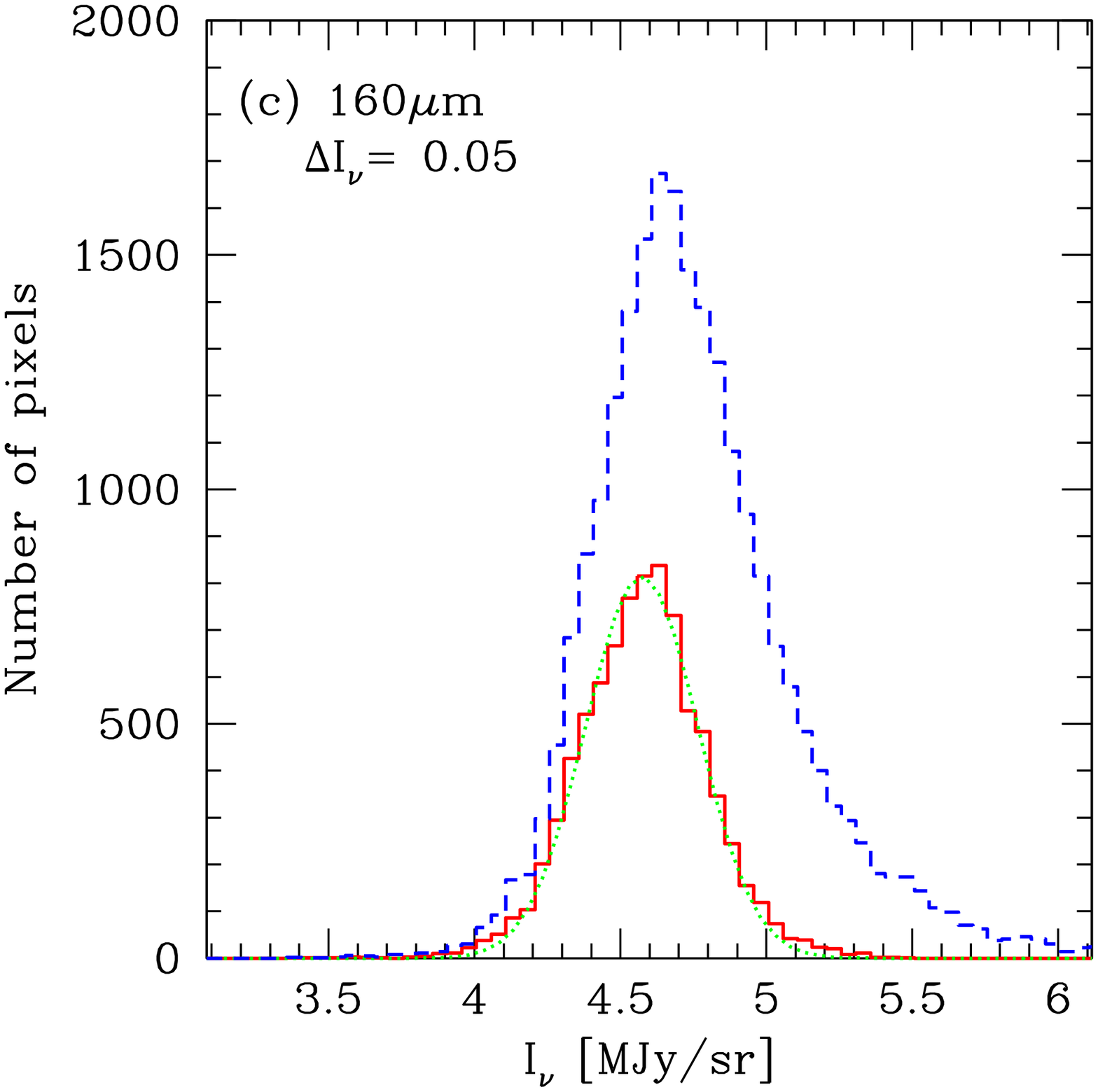}
\caption{Histograms of the image pixels before (dashed curves) and after
(solid curves) the source mask at (a) 24 $\mu$m, (b) 70 $\mu$m, and (c)
160 $\mu$m, respectively.  Dotted curves indicate, for reference, the
Gaussian distribution with $\sigma= 0.011$ MJy/sr, $0.11$ MJy/sr, and
$0.19$ MJy/sr at 24$\mu$m, 70$\mu$m, and 160$\mu$m, respectively.}
\label{fig:hist}
\end{figure}

Masking the regions around the detected sources removes 38\%, 66\% and
64\% of the pixels in the 24, 70 and 160 $\mu$m images, respectively. We
plot the histograms of the pixels with and without masking in Figure
\ref{fig:hist}. In all bands, the histograms before masking are
positively skewed due to bright sources.  Once the sources are excluded,
the data follow the Gaussian distribution more closely, while there are
small excesses at high $\sigma$. These excesses are due to
low-significance data affected by noisy detector arrays or low
coverage. Since the noise assigned to such data is large, their
contribution to the mean brightness is negligible.  At 160 $\mu$m, the
1$\sigma$ range of the distribution after the source mask, 0.19 MJy/sr,
is likely affected by the fluctuations of Galactic cirrus emission
\citep{Jeong05}. At 24 and 70 $\mu$m, the 1$\sigma$ ranges, $0.011$ and
$0.11$ MJy, are comparable to the average noise level listed in Table
\ref{tab:images}.

\begin{figure}
\epsscale{0.65} \plotone{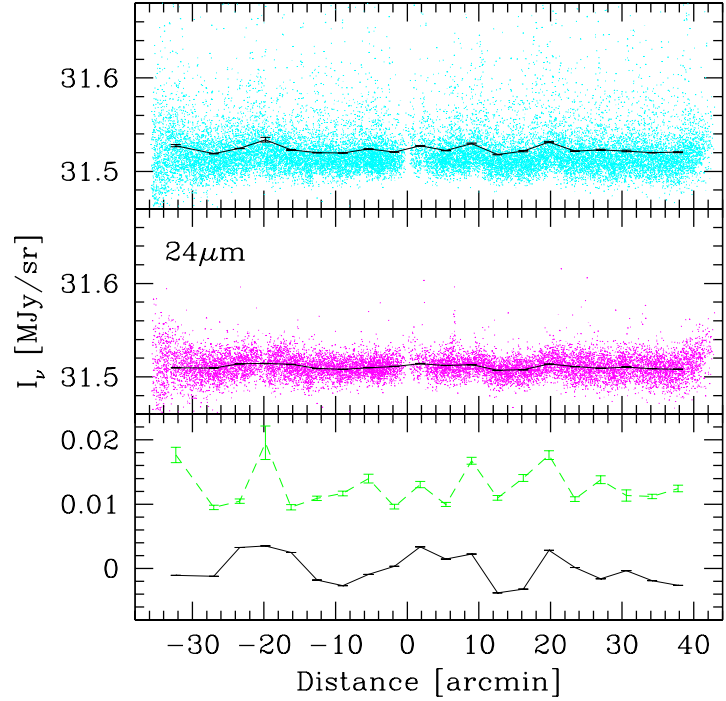} \caption{{\it Top:} distribution of the
24 $\mu$m brightness along the scan path before the source mask. For
clarity, we randomly select and display 1/10 of all the data
points. Solid curve indicates the mean brightness with the bin size of
$3.6'$ ($100$ kpc). {\it Middle:} same as the top panel except that the
surface brightness after the source mask is shown.  {\it Bottom:} dashed
curve shows the binned average of the removed component, i.e., the
difference between the mean brightness plotted in the top and middle
panels. For comparison, solid curve is the same as that in the middle
panel except that the average brightness of the whole map is subtracted
and the vertical scale is enlarged.  In all the panels, the distance is
measured from the X-ray center of the cluster with negative values
indicating the direction of decreasing declination. } \label{fig:prof1}
\end{figure}

\begin{figure}
\epsscale{0.65} \plotone{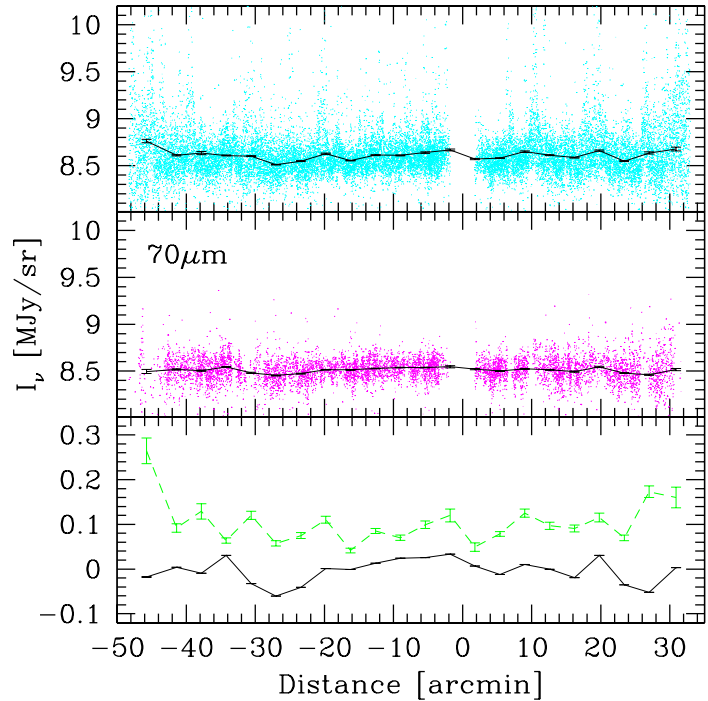} \caption{Same as Figure \ref{fig:prof1}
except that the 70 $\mu$m brightness is shown and, for clarity, 1/2 of
all the data points are displayed in the top and middle panels.  The 70
$\mu$m data are lacking within $1.7'$ from the X-ray center due to dead
pixels.}  \label{fig:prof2}
\end{figure}

\begin{figure}
\epsscale{0.65} \plotone{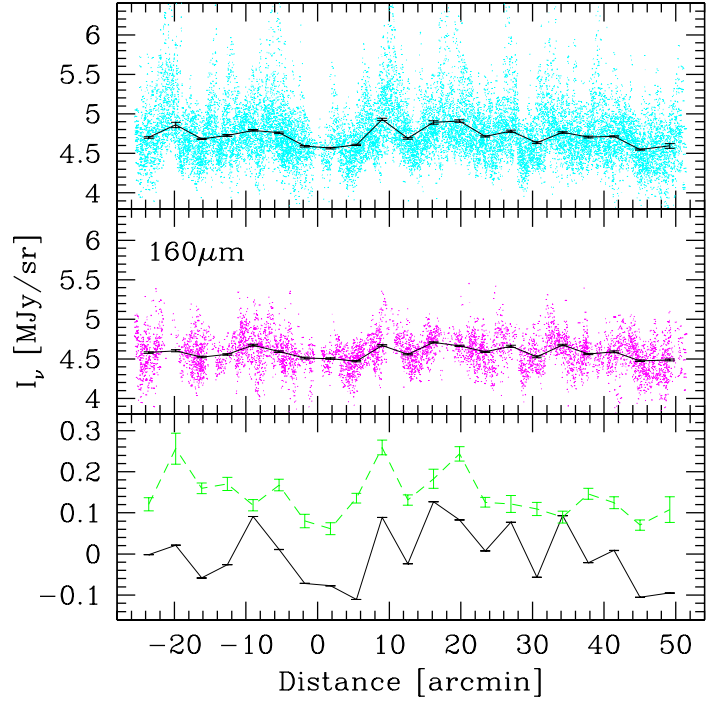} \caption{Same as Figure \ref{fig:prof1}
 except that the 160 $\mu$m brightness is shown and all the data points
 are displayed in the top and middle panels.}  \label{fig:prof3}
\end{figure}

Figures \ref{fig:prof1}--\ref{fig:prof3} show the surface brightness
distribution along the scan path as a function of the distance from the
X-ray center of the cluster (negative values of the distance are
assigned to the pixels at lower declination than the X-ray center). We
plot the brightness at each pixel and their binned average. Unless
otherwise stated, the error bar indicates the 1$\sigma$ statistical
error of the mean in each bin.  The 70 $\mu$m data are lacking within
$1.7'$ from the X-ray center due to dead pixels. We hence adopt
fiducially the bin size of $3.6'$ (100 kpc) in the present paper so that
it fully covers the deficit region while it is still sufficiently
smaller than the X-ray core radius of $10'$.  In all bands, the
fluctuations in the brightness, particularly the spikes due to point
sources, are significantly reduced after the source mask, yet there is
no indication of the central excess.

In the bottom panels of Figures \ref{fig:prof1}--\ref{fig:prof3}, we
plot the mean brightness of the removed component, i.e., the difference
between the unmasked and masked brightness.  From our linear scans
within $\sim 30'$ from the cluster center, we do not find a central
concentration of the integrated source intensities either.  The average
levels of the removed brightness are $0.012$ MJy/sr, $0.093$ MJy/sr and
0.16 MJy/sr at 24, 70 and 160 $\mu$m, respectively.

We further combine the brightness in the two directions from the cluster
center and compare its radial profile with model predictions in Figure
\ref{fig:prof_abs}. The prediction for the intracluster dust emission is
based on the theoretical model of \citet{Yamada05} (see the next section
for details).  For direct comparison with the {\it ISO} results, we have
adjusted the total amount of dust to reproduce the excess brightness of
0.2 MJy/sr at 120 $\mu$m within the radius of $1.5'$ inferred in
\citet{Stickel02}. Since we are only interested in the difference of the
signal across the cluster, an offset is added to the model brightness to
match the mean brightness of the whole map at the edge. At 24 $\mu$m,
the absence of the excess is as expected since the predicted temperature
of the intracluster dust is relatively low, $20-30$K (see also Dwek et
al. 1990). At 70 and 160 $\mu$m, the foreground/background fluctuations
are comparable to the expected levels of the emission and we do not find
a sign of the central excess in either band.

From the observed brightness within $r=3.6'$ relative to the mean
of the entire map, we obtain 2$\sigma$ upper limits on the central
excess as $5.2 \times10^{-3}$, $6.0 \times10^{-2}$ and $6.6\times
10^{-2}$ MJy/sr, at 24, 70 and 160 $\mu$m, respectively. In deriving
these limits, we have taken into account the background/foreground
fluctuations that dominate over the statistical noise as follows. We
have randomly selected a circular region of radius $3.6'$ from the map
in each band 100 times. From the variance of the mean brightness of
selected regions, we have assigned an $1\sigma$ error of $1.9 \times
10^{-3}$, $2.1 \times10^{-2}$ and $6.6\times 10^{-2}$ MJy/sr at 24, 70
and 160 $\mu$m, respectively.

\citet{Stickel98} and \citet{Stickel02} found with {\it ISO} the excess
far-infrared emission toward the center of this cluster based on the
surface brightness ratio $I_{120\mu\rm{m}}/I_{180\mu\rm{m}}$.  Although
the scan paths and the observed bands are not exactly the same in our
observations, we attempt to mimic their analysis by presenting
$I_{24\mu\rm{m}}/I_{70\mu\rm{m}}$ and $I_{70\mu\rm{m}}/I_{160\mu\rm{m}}$
in Figure \ref{fig:rat}.  While the fluctuation is large, there appears
to be a weak sign of a central enhancement in
$I_{70\mu\rm{m}}/I_{160\mu\rm{m}}$.  The enhancement is largely due to a
variation in the 160 $\mu$m map around the cluster center
(Fig.\ref{fig:images}); there are bright extended regions at $\sim 10'$
on each side, presumably produced by far-infrared sources lying on
fluctuating cirrus foreground. We have checked that no counterparts are
detected in the {\it Chandra} X-ray image of this region.  While some of
the sources are apparent and removed at 70 $\mu$m, the lower spatial
resolution and the larger cirrus fluctuation prevent us from removing
them completely at 160 $\mu$m. In turn, the cluster center appears to be
darker than its surrounding at 160 $\mu$m, resulting in the apparent
excess in $I_{70\mu\rm{m}}/I_{160\mu\rm{m}}$.  Figure \ref{fig:prof3}
also indicates that the depression at 160 $\mu$m is not due to
over-removal of the sources at the center. We therefore do not consider
this to be a detection of the intracluster dust.

\section{Discussion}

While there is no direct evidence of the central excess in any of the
three bands observed by the MIPS, we still find a weak sign of excess in the
brightness ratio $I_{70\mu\rm{m}}/I_{160\mu\rm{m}}$ that may resemble
the excess in $I_{120\mu\rm{m}}/I_{180\mu\rm{m}}$ reported by
\citet{Stickel98} and \citet{Stickel02}. It is therefore meaningful to
explore the consistency of these results.

For this purpose, we subtract from our data the average zodiacal light
and the cosmic infrared background levels estimated by the {\it Spitzer}
background estimator \citep{Reach00} at the scan center position for the
date of observation; 
$I_{\rm ZOD}= 7.21$ and $1.41$ MJy/sr, 
$I_{\rm CIB} = 0.21$ and $1.30$ MJy/sr at 70 and 160 $\mu$m,
respectively. We subtract the above values from the whole map since we
have already corrected the linear gradient of the map relative to the
scan center. We have checked that subtracting position-dependent
zodiacal light levels prior to the correction of the linear gradient
changes the resulting brightness ratio by no more than 5\%. The
uncertainties in the absolute calibration are not significant either;
$5\%$ at 70 $\mu$m and $12\%$ at 160 $\mu$m
\citep{Gordon07,Stansberry07}. However, simply because the mid-infrared
brightness is dominated by the zodiacal light, the nominal error of
$\sim 15\%$ in its model \citep{Reach00} will change the brightness
after subtraction at 70 and 160$\mu$m, $\sim 1.1$ MJy/sr and $\sim 1.9$
MJy/sr, by $\sim 100$\% and $\sim 10$\%, respectively.  Keeping these
large uncertainties in mind, we investigate the origin of the excess in
the band ratio as follows.

Figure \ref{fig:rat_abs} shows the radial distribution of the brightness
ratio $I^{\rm corr}_{70\mu\rm{m}}/I^{\rm corr}_{160\mu\rm{m}}$ after the
correction mentioned above. The error bars include only the statistical
uncertainties. Assuming that the corrected brightness distribution is
produced by a combination of a central excess and more extended
component, e.g., Galactic cirrus, we apply equation (7) of
\citet{Stickel02} to our observed bands and obtain $I_{70\mu\rm{m}}^{\rm
excess} \sim 6.4\times 10^{-2}$ MJy/sr within the radius of $3.6'$.
This value is comparable to the 2$\sigma$ upper limit of $6.0\times
10^{-2}$ MJy/sr derived solely from the 70 $\mu$m data in the previous
section.  The central excess, if real, should have been nearly canceled
at 70 $\mu$m by the fluctuation of the other component. The
interpretation of the central excess for the origin of the observed
$I_{70\mu\rm{m}}/I_{160\mu\rm{m}}$ profile is therefore rather unlikely,
although it is not fully rejected.

Note that equation (7) of \citet{Stickel02} is based on a key assumption
that the excess component contributes only to the variation of the
shorter wavelength data. Alternatively, it is also possible to interpret
the same data with the variation of the longer wavelength data.  In the
present observation, a plausible explanation may be that the corrected
brightness is still dominated by unresolved sources lying on the
fluctuating cirrus foreground, i.e., the absolute values of the signal
in both bands are dominated by such sources while the fluctuations at
160 $\mu$m are mainly produced by underlying cirrus emission.  In fact,
the value of $I^{\rm corr}_{70\mu\rm{m}}/I^{\rm corr}_{160\mu\rm{m}} =
0.55 \sim 0.6$ shown in Figure \ref{fig:rat_abs} corresponds to the dust
temperature of $T_d \sim 25$ K for the absorption efficiency of $Q_{\rm
abs} \propto \lambda^{-2}$ \citep{Draine84}. This temperature is higher
than the nominal temperature of Galactic cirrus $T_d\sim 18$ K for which
the brightness ratio is lower by more than a factor of 5.  The mean
values of $I^{\rm corr}_{70\mu\rm{m}} \sim 1.1$ MJy/sr and $I^{\rm
corr}_{160\mu\rm{m}} \sim 1.9$ MJy/sr are also higher than the levels of
the cirrus emission expected for this region, suggesting the
contribution of yet unresolved components.

\begin{figure}
\epsscale{0.65} \plotone{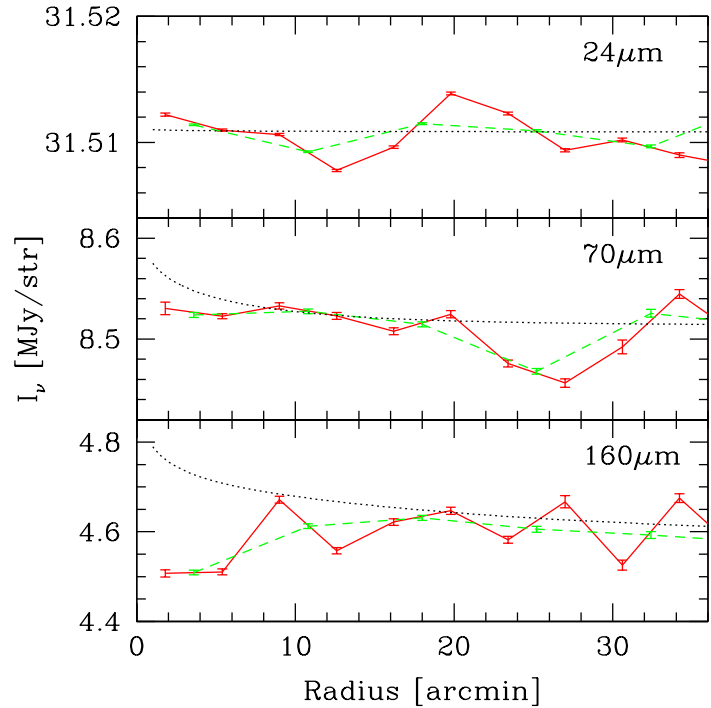} \caption{Radial profiles of the
surface brightness after the source mask at 24 $\mu$m ({\it top panel}), 70
$\mu$m ({\it middle panel}), and 160 $\mu$m ({\it bottom panel}), respectively.
Solid and dashed curves indicate the mean brightness with the bin size of
$3.6'$ and $7.2'$, respectively. For reference,
the prediction for the intracluster emission based the theoretical model
of \citet{Yamada05} is plotted in dotted curves (see the text for details).}
\label{fig:prof_abs}
\end{figure}

\begin{figure}
\epsscale{0.65} \plotone{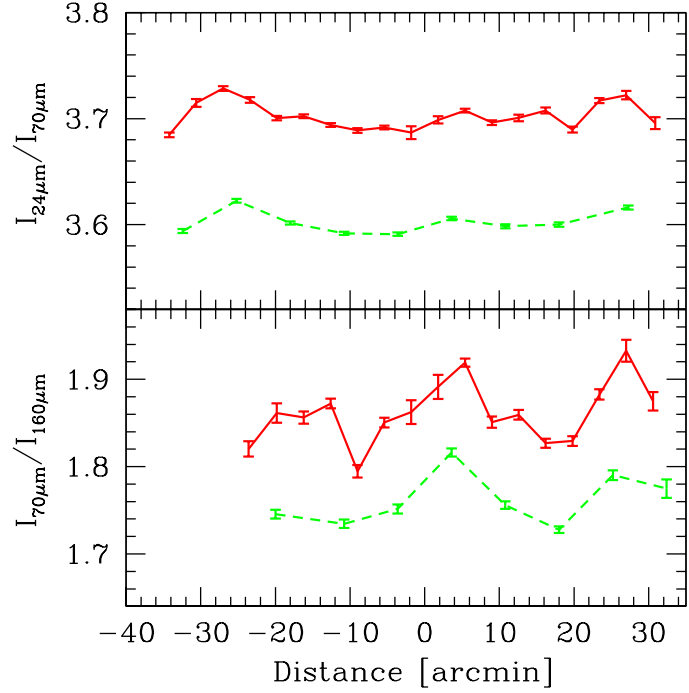} \caption{Distribution of the
brightness ratios $I_{24\mu\rm{m}}/I_{70\mu\rm{m}}$ ({\it top panel}) and
$I_{70\mu\rm{m}}/I_{160\mu\rm{m}}$ ({\it bottom panel}) after the source
mask along the scan path. The distance is measured from the X-ray center
of the cluster with negative values indicating the direction of
decreasing declination. Solid and dashed curves indicate the brightness
ratios with the bin size of $3.6'$ and $7.2'$,
respectively. For clarity, the dashed curves have been shifted
downwards by 0.1 in both panels.}  \label{fig:rat}
\end{figure}

\begin{figure}
\epsscale{0.65} \plotone{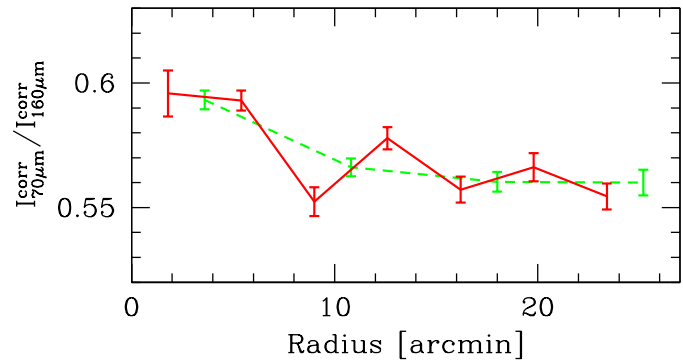} \caption{Radial profile of the
 brightness ratio $I^{\rm corr}_{70\mu\rm{m}} /I^{\rm
 corr}_{160\mu\rm{m}}$ after the source mask and the subtraction of the
 zodiacal light and the cosmic infrared background.  The error bars
 include only the statistical uncertainties. Solid and dashed curves
 indicate the brightness ratios with the bin size of $3.6'$ and $7.2'$,
 respectively.}  \label{fig:rat_abs}
\end{figure}

From the limits on the infrared brightness in each band, we further
attempt to derive constraints on the amount of the intracluster dust.
Model predictions shown in Figure \ref{fig:prof_abs} \citep{Yamada05}
rely on an assumption that the dust is supplied continuously to the
intracluster medium from galaxies with the initial size distribution
similar to our Galaxy, $dn^{\rm init}/da \propto a^{-3.5}$
\citep{Mathis77}, and has reached a steady state against sputtering with
the size distribution $dn/da \propto a^{-2.5}$. The galaxy distribution
is assumed to trace that of the dark matter given by the Navarro--Frenk
--White density profile \citep{NFW97}, and the gas density follows the
isothermal beta model.  Given these, the curves plotted in Figure
\ref{fig:prof_abs} correspond to the dust surface density of $\Sigma_d =
1.1 \times 10^3 $ M$_\odot$ kpc$^{-2}$ within the central $3.6'$ of the
cluster.  The 2$\sigma$ upper limits on the excess brightness derived in
the previous section are 1.3 and 0.37 times the model brightness
averaged within the central $3.6'$ at $70\mu$m and $160\mu$m,
respectively.  Considering much lower uniformity of the data at 160
$\mu$m, we take the limit of the $70\mu$m data to obtain $\Sigma_d < 1.4
\times 10^{3}$ M$_\odot$ kpc$^{-2}$. This limit corresponds to the
dust-to-gas-mass ratio of less than $1 \times 10^{-5}$ within the
central 100 kpc, which is nearly 3 orders of magnitude smaller than the
Galactic value \citep{Draine03}.

The average visual extinction $A_V$ is related to the dust surface density 
as \citep{Spitzer78}
\begin{eqnarray}
A_V &=& 0.011 \left( \frac{Q_{\rm ext}}{2}\right) 
\left(\frac{\Sigma_d}{10^{3} \mbox{ M}_\odot \mbox{kpc}^{-2}} \right)
\nonumber \\
&& \times \left(\frac{\rho_d}{3 \mbox{ g cm}^{-3}} \right)^{-1}
\left(\frac{a}{0.1 \mu \mbox{m}}\right)^{-1}
\end{eqnarray}
where $Q_{\rm ext}$ is the extinction efficiency factor, $\rho_d$ is the
dust mass density, and $a$ is the grain radius. Adopting the standard
values for $Q_{\rm ext}$, $\rho_d$ and $a$ quoted in the above equation,
our derived limit of $\Sigma_d < 1.4 \times 10^{3}$ M$_\odot$ kpc$^{-2}$
corresponds to $A_V < 0.016$ within the central 100 kpc. On the other
hand, \citet{Chelouche07} report the detection of the mean $E(B-V)$ of a
few times $10^{-3}$ mag toward SDSS QSOs behind $\sim 10^4$ clusters,
and \citet{Muller08} obtain the limit $\langle A_V \rangle < 0.024$
(2$\sigma$) from reddening of $\sim90,000$ background galaxies for 458
clusters. While these optical measurements are given for much larger
($>$ Mpc) scales, our result is still consistent with them.

Finally, gravitational lensing can impact on the emission of background
sources behind galaxy clusters. As the total surface brightness is
conserved by lensing, the bias in the resolved or unresolved brightness
is caused solely by the sources that exceed the detection limit as a
result of magnification \citep{Refregier97}.  The number of such sources
is $ f \equiv \mu^\alpha-1$ times the number of unlensed sources already
lying above the flux limit $S$, where $\mu$ is the magnification and
$\alpha$ is the index of the unlensed source count, $N(>S)\propto
S^{-\alpha}$.  The fractional change in the resolved brightness has the
same sign as and is less than $f$, because the sources contributing to
the change have the lowest flux among those resolved. Note that the
factor $f$ is different from the bias in the overall surface density
given by $\mu^{\alpha-1}$ \citep[e.g.,][]{Bartel01}.  Recent surveys
with {\it Spitzer} yield the mean values of $\alpha = 0.5$ down to 35
$\mu$Jy at 24 $\mu$m \citep{Papovich04} and $\alpha=0.6$ down to 1.2 mJy
at 70 $\mu$m \citep{Frayer06}. The depths of these measurements are
comparable to those of the present observation. At 160 $\mu$m, the
published counts are currently available for $S > 50$ mJy, and an
extrapolation using the model of \citet{Lagache04} yields $\alpha \sim
1.5$ at $S \sim 20$ mJy \citep{Dole04a}.  In the case of Coma, the
Einstein radius is $\sim 30''$ and the magnification is expected to vary
from $\mu \sim 1.2$ at $r=3.6'$ to $\mu \sim 1.02$ at $r=30'$ for a
singular isothermal mass profile. Taken together, the above factor in
this radius range varies from $f \sim 0.1$ to $0.01$ at 24 $\mu$m and 70
$\mu$m, and $f \sim 0.3$ to $0.03$ at 160 $\mu$m. The fractional change
in the brightness of the {\it removed component} shown in Figures 4--6
should therefore be smaller than the variations along the scan
path. There is in fact no sign of systematic increase in the resolved
brightness toward the cluster center in any band. Owing to the
conservation of the total brightness, the bias in the residual
unresolved background should also be small.  We hence consider that the
impact of gravitational lensing is negligible in the present results.

\section{Conclusions}

In this paper, we have undertaken a search for the infrared emission
from the intracluster dust in the Coma cluster of galaxies by the MIPS
on board the {\it Spitzer} Space Telescope. Our observations yield the
deepest mid and far-infrared images of a galaxy cluster ever achieved
with average 1$\sigma$ sensitivities per pixel of $1.3\times10^{-2}$
MJy/sr, $1.1\times 10^{-1}$ MJy/sr and $1.3\times 10^{-1}$ MJy/sr at 24,
70 and 160 $\mu$m, respectively. In each of the three bands, we have not
detected a signature of the central excess component. Our observations
yield the 2$\sigma$ upper limits on the excess emission within $3.6'$
(100 kpc) of the cluster center as $5.2 \times10^{-3}$ MJy/sr, $6.0
\times10^{-2}$ MJy/sr, and $6.6\times 10^{-2}$ MJy/sr, at 24, 70, and
160 $\mu$m, respectively.  These results are consistent with the values
reported by \citet{Bai07} for Abell 2029, as well as those derived from
a stacking analysis of 11,507 clusters \citep{Montier05}.

With the improved spatial resolution and sensitivity of the MIPS, we
have obtained much more robust limits on the excess emission from this
cluster than previous observations. We still find that the brightness
ratio between 70$\mu$m and 160$\mu$m shows a marginal sign of the
central excess, in qualitative agreement with the finding of
\citet{Stickel98} and \citet{Stickel02} by {\it ISO}. Although we cannot
fully reject the contribution of the intracluster dust, it is more
likely due to unresolved infrared sources lying on fluctuating cirrus
foreground.

Combined with the theoretical model of \citet{Yamada05}, we have derived
a limit on the surface mass density of the intracluster dust as
$\Sigma_d < 1.4 \times 10^{3}$ M$_\odot$ kpc$^{-2}$. This limit
corresponds to the dust-to-gas-mass ratio of less than $10^{-5}$ within
the central 100 kpc, which is nearly three orders of magnitude
smaller than the Galactic value, implying that the dust destruction is
very efficient at the center of the cluster. Our result can further be
converted to the limit on the visual extinction of $A_V < 0.02$, which
is consistent with the results derived from statistical reddening of
background objects behind a large sample of galaxy clusters
\citep{Chelouche07, Muller08}.

\bigskip
\bigskip

We thank Takashi Hamana, Nobuhiro Okabe and Norio Sekiya for discussions
and an anonymous referee for helpful comments.  This work is supported
in part by Grant-in-Aid for Young Scientists by MEXT (18740112,
19740112). N.O. acknowledges support from the Alexander von Humboldt
Foundation.


\begin{table}
\caption{Parameters of the images}
\label{tab:images}
\begin{center}
\begin{tabular}{cccc}
\hline \\[-9pt] 
\hline \\[-6pt] 
Wavelength & 24 $\mu$m & 70 $\mu$m & 160 $\mu$m \\[4pt]\hline \\[-6pt]
Pixel size [$''$] & 2.45 & 4.0 & 8.0 \\
FWHM [$''$] & 6.2 & 18 & 40 \\
Average noise per pixel [MJy/sr] &$1.3\times
10^{-2}$  & $1.1\times 10^{-1}$  & $1.3\times 10^{-1}$ \\[4pt] \hline 
\end{tabular} 
\end{center}
\end{table}


\end{document}